\documentclass[aps,prb,twocolumn,superscriptaddress,floatfix,nofootinbib]{revtex4-2}

\usepackage{amsmath,amssymb}
\usepackage{bm}
\usepackage{graphicx}

\newcommand{\Tr}{\mathrm{Tr}}

\begin{document}

\title{Role of chirality in the polarizer-filter dichotomy in nonmagnetic junctions}

\author{S. Urazhdin}
\affiliation{Department of Physics, Emory University, Atlanta, Georgia 30322, USA}

\date{\today}

\begin{abstract}
Chirality-induced spin selectivity can enable spin generation by nonmagnetic junctions, but its symmetry requirements remain debated. We show that the symmetry of the junction constrains its polarizing behavior: a spin polarizer requires a chiral junction, whereas an achiral junction can act as a spin filter. This symmetry involves the junction as a whole, not only the molecule. Achiral molecules can act as polarizers or filters depending on junction geometry, as confirmed with a minimal tight-binding model. Reciprocity forbids polarization detection by linear two-terminal magnetoresistance; we propose a nonlocal measurement that distinguishes polarizers from filters and establishes whether the junction breaks time-reversal symmetry.
\end{abstract}

\maketitle

\paragraph*{Introduction.} Electrons transmitted through chiral molecules can become spin polarized along their propagation direction, with the sign set by the molecular handedness~\cite{Ray1999,Gohler2011,NaamanNRC2019,Bloom2024,Evers2022}. This chirality-induced spin selectivity (CISS) is manifested in junctions of chiral molecules with ferromagnets (FMs) as a two-terminal (2T) magnetoresistance (MR) odd in both the magnetization $\bm M$ and the chirality~\cite{Bloom2024}. The MR is commonly interpreted by analogy with a spin valve, in which the molecule polarizes the current and the FM analyzes it, or vice versa~\cite{NaamanNRC2019, Bloom2024}.

Unlike a FM, a nonmagnetic molecule preserves time-reversal symmetry. For a junction with a single FM, the Onsager--Casimir relation then gives $G(\bm M)=G(-\bm M)$, forbidding linear 2T MR~\cite{Onsager1931,Casimir1945,Buttiker1986}. Nevertheless, CISS MR is commonly reported at small bias~\cite{Liu2020,Hedegard2026}, spurring an ongoing debate about the spin-polarizing properties of nonmagnetic junctions~\cite{Evers2022,Bloom2024,Yan2024,GarciaBlazquez2023,Rikken2023,Tirion2024,Sala2026}. 

Three known requirements must be satisfied for a nonmagnetic 2T junction to polarize spin. First, time reversal symmetry prohibits polarization of electrons scattered into a lead unless this lead carries multiple orbital channels~\cite{ZhaiXu2005,Bardarson2008,Utsumi2020}. Second, to produce polarization along the transport axis, the junction cannot have a mirror symmetry plane $\Sigma_l$ containing this axis~\cite{Guo2016,Dednam2023}. A third requirement, derived or assumed necessary for chiral junctions, is spin flipping of reflected electrons~\cite{Utsumi2020,YangVanDerWal2019,Tirion2024}. 
Molecular chirality is sufficient to eliminate $\Sigma_l$, but not necessary. The general relation between the junction symmetry and its polarizing properties remains to be established.

\begin{figure}[t]
	\centering
	\includegraphics[width=\columnwidth]{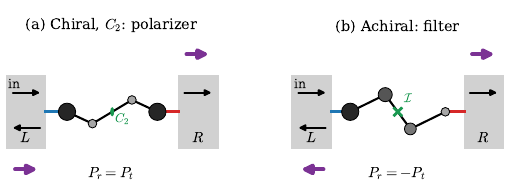}
	\caption{(a) A twisted chain with a $C_2$ axis normal to the page forms a chiral junction polarizing transmitted and reflected electrons (purple arrows) in the same direction: it is a polarizer. (b)  A tilted planar chain with an inversion center ($\times$) forms an achiral junction polarizing transmitted and reflected electrons oppositely: it is a filter. Larger atoms are closer to the viewer.\label{fig:schematic}}
\end{figure}

A central spin-transport characteristic of a 2T junction is the relation between the polarizations in its two leads. A junction can be defined as a spin polarizer if transmitted and reflected electrons are polarized in the same direction, and a spin filter if it reflects the spin it does not transmit, as exemplified by FMs~\cite{Wolf2022}, Fig.~\ref{fig:schematic}. Nonmagnetic junctions symmetric under a $\pi$ rotation exchanging the leads cannot act as filters~\cite{Utsumi2020,Wolf2022}. This conclusion was extended to all junctions by assuming that the transmitted polarization reverses with the transport direction~\cite{Wolf2022,Yan2024}, but the general conditions for its validity have not yet been established. Since FMs act as filters, the class of a chiral junction determines whether the analogy with a FM in spin-valve interpretations of CISS MR is applicable.

Here we establish the general relation between the symmetry of a junction and its polarizing class. A proper lead-exchanging symmetry allows only polarizer behavior, and an improper one only filter behavior. A polarizer therefore requires a chiral junction, but not necessarily a chiral molecule. For elastic scattering, both classes require spin-flip reflection, generalizing results obtained for special cases~\cite{ZhaiXu2005,Bardarson2008,Utsumi2020,YangVanDerWal2019,Tirion2024}. Unlike FMs, both classes preferentially transmit in one direction the spin opposite to that they generate in the other. We illustrate these results with a minimal model and propose a nonlocal measurement that distinguishes the two classes.

\paragraph*{Scattering relations.} We consider elastic scattering by a nonmagnetic junction between leads $L$ and $R$, which connect the junction to metallic electrodes acting as charge reservoirs. The lead channels are assumed to have negligible spin-orbit coupling (SOC) or inversion symmetry, and are consequently spin degenerate. The scattering matrix is $S=\begin{pmatrix} r & t'\\ t & r'\end{pmatrix}$, with $r$, $t$ ($r'$, $t'$) the reflection and transmission blocks for electrons incident from $L$ ($R$). Each channel is indexed by the lead, the orbital mode and the spin projection on the transport axis $\hat{\bm z}$. For unpolarized incident electrons, the differences between the probabilities of transmission (reflection) into spin-up and spin-down channels are
\begin{equation}
\begin{aligned}
P_t&=\Tr[\sigma_z tt^\dagger], & P_r&=\Tr[\sigma_z rr^\dagger],\\ P_{t'}&=\Tr[\sigma_z t't'^\dagger], & P_{r'}&=\Tr[\sigma_z r'r'^\dagger].
\end{aligned}
\label{eq:defs}
\end{equation}
For brevity, we refer to them as polarizations. The junction is a polarizer if $P_tP_r>0$ and a filter if $P_tP_r<0$.

Unitarity and the spin degeneracy of the lead channels give~\cite{KiselevKim2005,Wolf2022}
\begin{equation}
P_t+P_{r'}=0,\qquad P_{t'}+P_r=0,
\label{eq:sumrule}
\end{equation}
i.e., the spin current in each lead vanishes in equilibrium. Since $P_tP_r=P_{t'}P_{r'}$, the class does not depend on the side from which the electrons are incident.

Time reversal gives $S=\mathcal{Z}S^T\mathcal{Z}^\dagger$, with $\mathcal{Z}=U\otimes i\sigma_y$, where $U$ maps each orbital mode onto its time-reversed partner~\cite{Beenakker1997,Bardarson2008}. Since $\mathcal{Z}^\dagger\sigma_z\mathcal{Z}=-\sigma_z$, this relation constrains the reflection into each lead. For a single-channel lead, the reflection block is a self-dual $2\times2$ matrix, $r=\sigma_yr^T\sigma_y$, giving $r\propto1$: the electrons emitted into this lead are unpolarized~\cite{ZhaiXu2005,Bardarson2008,Utsumi2020}. 

A similar constraint was demonstrated for spin-conserving reflection into  multichannel leads by spin polarizers based on chiral molecules~\cite{Utsumi2020,YangVanDerWal2019,Tirion2024}. We generalize this result to both junction classes. Time reversal maps the spin-up block of the spin-conserving reflection onto the transpose of the spin-down block, so both spins are reflected with the same total probability. For lead $L$, this gives $P_r=0$ and, by Eq.~(\ref{eq:sumrule}), $P_{t'}=0$ (SM~S1). 
Similarly, spin-conserving reflection into $R$ gives $P_{r'}=P_t=0$. Thus, the electrons emitted into a lead are polarized only if the reflection into this lead includes spin flip, so both polarizers and filters require spin-flip reflection into both leads.

Next, we consider the effect of junction symmetry. A symmetry operation $g$ acts on $S$ through a unitary $\mathcal{G}$, $S=\mathcal{G}S\mathcal{G}^\dagger$. A mirror plane $\Sigma_l$ preserves the leads and reverses the spin along $\hat{\bm z}$, giving $P_t=P_r=0$~\cite{Dednam2023}. An operation $g$ exchanging the leads maps $t$ onto $t'$ and $r$ onto $r'$. It also multiplies $\sigma_z$ by $\varepsilon_g=-\det g$, with $\varepsilon_g=-1$ for proper and $\varepsilon_g=1$ for improper $g$, since spin is an axial vector. Thus $P_{t'}=\varepsilon_gP_t$ and $P_{r'}=\varepsilon_gP_r$. With Eq.~(\ref{eq:sumrule}),
\begin{equation}
P_r=-\varepsilon_gP_t.
\label{eq:class}
\end{equation}
A proper lead-exchanging symmetry gives $P_r=P_t$: the junction can only be a polarizer, and its transmitted polarization reverses with the transport direction. An improper one gives $P_r=-P_t$: the junction can only be a filter, and its transmitted polarization does not depend on the transport direction. The relation between $P_t$ and $P_r$ for a filter is the same as for a FM. However, unlike a FM, both classes preferentially transmit from one lead the spin opposite to that they produce in transmission from the other, as shown below.

\begin{table}[b]
\caption{Constraints imposed by spatial symmetry on the polarizing properties of nonmagnetic 2T junctions; $g$ is a lead-exchanging symmetry, $\Sigma_l$ is a mirror plane containing the transport axis.\label{tab:classes}}
\begin{ruledtabular}
\begin{tabular}{lll}
Junction symmetry & Chirality & Constraint \\
\hline
$\Sigma_l$ & achiral & $P_t=P_r=0$\\
proper $g$, no $\Sigma_l$ & chiral & $P_r=P_t$\\
improper $g$, no $\Sigma_l$ & achiral & $P_r=-P_t$\\
no $g$, no $\Sigma_l$ & chiral & none\\
\end{tabular}
\end{ruledtabular}
\end{table}

Table~\ref{tab:classes} summarizes the classification. An achiral junction is symmetric under an improper operation. If this operation preserves the leads, it is $\Sigma_l$, and the polarization vanishes. Otherwise, the junction can only be a filter. Junctions without lead-exchanging symmetry are unconstrained and can act as polarizers or filters.

\paragraph*{Minimal model.} We illustrate the classification for a chain of atoms carrying three $p$ orbitals, with orbitally-selective nearest-neighbor hopping described by the Slater--Koster parameters $V_{pp\sigma}=1$ serving as the energy scale  and $V_{pp\pi}=-0.4$~\cite{SlaterKoster1954}. The atomic SOC energy is $\lambda\bm L\cdot\bm S$ with $\lambda=0.4$, the dependence of the polarizing properties on $\lambda$ is discussed in SM~S2. We neglect the contact contribution, which does not affect the symmetries of the polarizing properties.  Each lead has three orbital channels selectively coupling to the $p$ orbitals of an end atom, with amplitude $1$ for $p_z$ and $0.5$ for $p_x$, $p_y$; this defines leads axially symmetric with respect to $\hat{\bm z}$. The symmetry of the junction is set by the atomic positions and the lead directions (SM~S2).

\begin{figure}[t]
\centering
\includegraphics[width=\columnwidth]{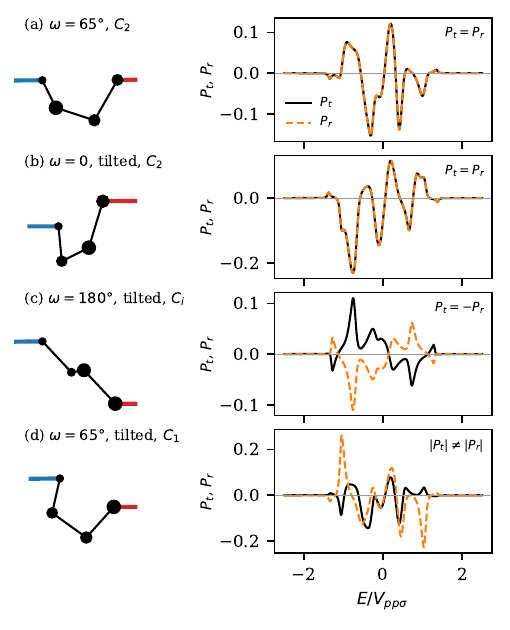}
\caption{Transmitted polarization $P_t$ (solid) and reflected polarization $P_r$ (dashed) vs energy $E$ for a four-atom chain between leads oriented along $\hat{\bm z}$ (blue: source, red: drain), for configurations with different dihedral angles $\omega$, schematically shown on the left. (a),(b) Polarizers: chiral junctions with a lead-exchanging $C_2$ axis, $\omega=65^\circ$ (a) and $\omega=0$ tilted about the $C_2$ axis (b). (c) Filter: achiral junction with an inversion center, $\omega=180^\circ$, tilted so that the molecular plane does not contain $\hat{\bm z}$. (d) Junction without symmetry, $\omega=65^\circ$ with a generic tilt: polarizer or filter depending on $E$. Larger atoms are closer to the viewer.\label{fig:classes}}
\end{figure}

A four-atom chain with bond angle $110^\circ$ realizes all classes of Table~\ref{tab:classes} by changing the orientation relative to the leads and the dihedral angle $\omega$, the angle between the planes of the first three and the last three atoms. The energy $E$ is measured from the on-site energy of the $p$ orbitals and corresponds to the equilibrium Fermi energy of the leads. 

For $\omega=0$ and $180^\circ$, the chain is planar. If it is untilted, its plane contains $\hat{\bm z}$ and is a mirror plane $\Sigma_l$ of the junction, so $P_t=P_r=0$. The twisted chain with $\omega=65^\circ$ is chiral, and its only symmetry is a $C_2$ axis perpendicular to $\hat{\bm z}$, giving a polarizer, Fig.~\ref{fig:classes}(a). The achiral chain with $\omega=0$, tilted about its $C_2$ axis, forms a chiral junction with the same behavior, Fig.~\ref{fig:classes}(b). The chain with $\omega=180^\circ$, tilted so that its plane does not contain $\hat{\bm z}$, retains only the inversion center, giving a filter, Fig.~\ref{fig:classes}(c). This confirms that achiral molecules can act as polarizers or filters, depending on the geometry of the whole junction. Without any symmetries, $P_r/P_t$ varies with energy, Fig.~\ref{fig:classes}(d). In this case, the class depends on energy, structure, the coupling to the leads, and the SOC strength (SM~S2).

\paragraph*{Two-terminal magnetoresistance.} Our model also illustrates how reciprocity prevents the detection of the polarization by 2T MR~\cite{YangVanDerWal2019,Tirion2024}. We replace lead $R$ by a FM with spin polarization $p$ along $\bm M\parallel\pm\hat{\bm z}$. In the spin-valve picture, the junction polarizes the electrons and the FM analyzes them, giving a magnetization-odd transmission $\propto pP_t$. However, electrons reflected by the FM are reflected back by the junction with spin-dependent spin flip, and the multiple reflections cancel this contribution. For the junction of Fig.~\ref{fig:classes}(a) with $p=0.5$, $G(\bm M)=G(-\bm M)$ to numerical precision, while $|P_t|$ reaches $0.15$ (SM~S2, Fig.~S3). 
Time reversal relates the two transport directions. The spin asymmetry of transmission from $R$ is $A'= \Tr[\sigma_z t'^\dagger t']=-P_t$ (SM~S1). Thus, unlike a FM, the junction preferentially transmits from $R$ the spin opposite to that it produces in transmission from $L$. The spin-valve contributions of the two transport directions therefore have opposite signs, and the linear 2T MR vanishes.

\begin{figure}[t]
	\centering
	\includegraphics[width=\columnwidth]{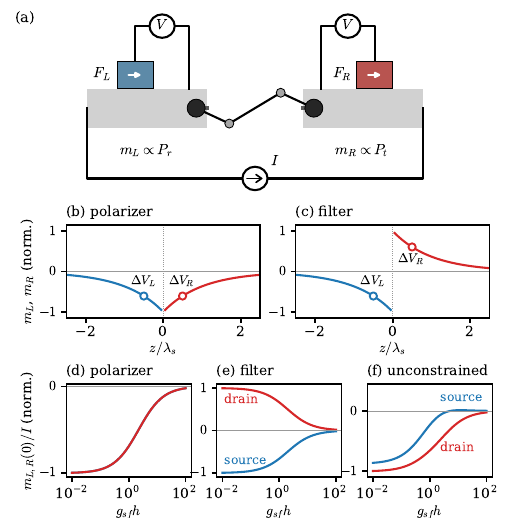}
	\caption{(a) Nonlocal detection: the current $I$ produces spin accumulations $m_{L,R}$ in the electrodes, separately detected by FM probes $F_{L,R}$. (b),(c) $m_L$ ($z<0$) and $m_R$ ($z>0$) vs position for the polarizer and the filter of (d),(e) at $g_{sf}h=10$. Circles: $|z|=\lambda_s/2$. (d)--(f) Accumulations at the junction $m_{L,R}(0)$ in the source (blue) and drain (red) electrodes vs spin-flip conductance $g_{sf}$ for identical electrodes, normalized to the maximum of $|m_{L,R}(0)/I|$: (d) polarizer of Fig.~\ref{fig:classes}(a), (e) filter of Fig.~\ref{fig:classes}(c), (f) unconstrained junction of Fig.~\ref{fig:classes}(d). Panels (b)--(f) are calculated with the circuit model including the spin backflow (SM~S3). All calculations use $E_F=-0.3$. Metallic electrodes correspond to $g_{sf}h\sim10^4$, where the behavior is the same as at the largest $g_{sf}$ shown.
	\label{fig:detection}}
\end{figure}

\paragraph*{Nonlocal detection.} The proposed nonlocal geometry includes FM probes attached to the electrodes on the two sides of the junction (Fig.~\ref{fig:detection}(a)), similar to nonlocal CISS geometry already implemented with nonmagnetic electrodes~\cite{Liu2024}. The polarized electrons injected into the electrodes produce spin accumulations $m_{L,R}(z)=\mu^\uparrow_{L,R}(z)-\mu^\downarrow_{L,R}(z)$, with $z$ the position relative to the junction. Each accumulation is detected as a voltage between the FM probe and the lead odd in magnetization $\bm M$, $\Delta V_{L,R}=V(\bm M)-V(-\bm M)\propto m_{L,R}$~\cite{JohnsonSilsbee1985,Jedema2001}. Unlike the 2T conductance, this signal is not forced to vanish by reciprocity. Multiterminal reciprocity, $R_{ij,kl}(\bm M)=R_{kl,ij}(-\bm M)$~\cite{Buttiker1986}, only relates it to the reciprocal measurement. 

In the reciprocal measurement, bias between lead $R$ and the FM probe attached to it produces a spin accumulation in $R$, raising the chemical potential of spin-up electrons and lowering that of spin-down electrons. Spin-up electrons then flow from $R$ to $L$, and electrons from $L$ flow into the empty spin-down states of $R$. The net current from $R$ to $L$ is proportional to the spin asymmetry of transmission from $R$, given by the transmission spin asymmetry $A'=\Tr[\sigma_zt'^\dagger t']$. Time reversal gives $A'=-P_t$ (SM~S1). Thus, for $P_t>0$, if a current from $L$ to $R$ produces a spin-up accumulation in $R$, the same  spin-up accumulation imposed in $R$ drives a current from $L$ to $R$, as required by reciprocity. For a FM, $A'=P_t$, and the imposed accumulation drives the current in the opposite direction. Comparing the direct and reciprocal signals measured with the same probe therefore tests whether the junction is nonmagnetic.

The relative sign of $m_L$ and $m_R$ follows from the lead-exchanging symmetry for any transport mechanism. The operation $g$ exchanges the electrodes, reverses the current $\bm J$, and multiplies the accumulation by $\varepsilon_g$. Thus $m_R(z,\bm J)=\varepsilon_g m_L(-z,-\bm J)$, and the linear response satisfies $m_L(-z)/m_R(z)=-\varepsilon_g$. The response even in $\bm J$ satisfies the opposite relation and is removed by reversing the current (SM~S3). For a polarizer, $\varepsilon_g=-1$ and the linear response is even in $z$; for a filter, $\varepsilon_g=1$ and it is odd, Figs.~\ref{fig:detection}(b),(c).

For elastic transport, Eq.~(\ref{eq:sumrule}) gives the spin currents into the unpolarized electrodes,
\begin{equation}
	I^s_R=\frac{P_t}{h}(\mu_L-\mu_R),\qquad I^s_L=\frac{P_r}{h}(\mu_L-\mu_R),
	\label{eq:Is}
\end{equation}
with $\mu_{L,R}$ the chemical potentials of the leads.  The accumulation also drives a spin current back through the junction. This backflow is negligible, since the spin resistance $R_s=\rho\lambda_s/A$ of a metallic electrode with resistivity $\rho$, spin diffusion length $\lambda_s$, and cross section $A$ is much smaller than the junction resistance~\cite{Brataas2000} (SM~S3). The spin injected into each electrode is then balanced by its relaxation,
\begin{equation}
	I^s_{L,R}=g_{sf}\,m_{L,R}(0),
	\label{eq:balance}
\end{equation}
where $g_{sf}=1/(2e^2R_s)$ is the spin-flip conductance of the electrode. For identical electrodes, $m_L(0)/m_R(0)=P_r/P_t=-\varepsilon_g$, and $m_{L,R}(z)=m_{L,R}(0)e^{-|z|/\lambda_s}$. For equal $g_{sf}$ in the two electrodes, the full calculation including the backflow (SM~S3) gives the same ratios at all $g_{sf}$, Figs.~\ref{fig:detection}(b)--(e). For the unconstrained junction, the backflow can reverse the relative sign of the accumulations if $e^2g_{sf}\lesssim G$, Fig.~\ref{fig:detection}(f). For metallic electrodes, $e^2g_{sf}\gg G$, and the relative sign is that of $P_r/P_t$ (SM~S3).

These relations define the measurement protocol. Reversing the probe magnetizations removes voltages unrelated to the spin accumulation, such as the ohmic voltage along the electrodes. Reversing the current isolates the response linear in $\bm J$, removing contributions even in $\bm J$, such as thermoelectric voltages. The junction class is identified by the relative sign of the signals measured by the two probes: the same sign for a polarizer, and opposite signs for a filter.

We now estimate the expected signals. A spin current $PI/e$ injected into an electrode, with $P=P_t/\Tr[tt^\dagger]$ the current polarization, gives $\Delta V\approx2p_DPR_sIe^{-L/\lambda_s}$ at a probe with polarization $p_D$ at a distance $L$~\cite{JohnsonSilsbee1985,Jedema2001,TakahashiMaekawa2003} (SM~S3). For Cu electrodes with $A=100\times50$~nm$^2$, $R_s\approx1.4$--$2.8$~$\Omega$ at room and low temperature~\cite{Jedema2001}. With $P=0.1$, $p_D=0.3$, $L=\lambda_s/2$, and $I=0.1$--$1$~$\mu$A, $\Delta V\approx5$--$100$~nV.

\paragraph*{Discussion.} Our main finding is that the junction symmetry with respect to lead exchange constrains whether it acts as a polarizer or a filter. For elastic scattering, both classes require spin-flip reflection, generalizing the earlier results for special cases~\cite{ZhaiXu2005,Bardarson2008,Utsumi2020,YangVanDerWal2019}. Only the polarizer requires chirality, which is a property of the junction as a whole and does not require molecular chirality. Achiral molecules tilted relative to the transport axis, as is common in self-assembled monolayers~\cite{Guo2016,Dednam2023}, can act as polarizers or filters. A testable prediction is CISS-like MR in junctions of achiral molecules with controlled tilt sense. For isotropic ensembles, chiral molecules act as polarizers while the polarization produced by achiral molecules averages to zero (SM~S2), consistent with the polarizer behavior concluded in Refs.~\cite{Wolf2022,Yan2024} (SM~S2). 

Discriminating between polarizer and filter requires measuring the polarization on both sides of the junction. Photoemission through chiral monolayers measures only the transmitted electrons~\cite{Ray1999,Gohler2011}, and cannot determine the class. Optical imaging of C$_2$-symmetric chiral nanowire-based junctions confirmed polarizer behavior~\cite{Lee2025}, consistent with  Table~\ref{tab:classes}. An extension of such measurements to junctions lacking this symmetry can test the presented analysis. The proposed electrical nonlocal measurement also applies to molecular junctions whose contacts are not optically accessible. The signal establishes the polarization, and the relative sign of the two signals establishes the class. This tests whether junctions exhibit polarizer behavior, as assumed in Refs.~\cite{YangVanDerWal2019,YangVanDerWal2020}, or filter behavior, as implied by the analogy with a FM underlying spin-valve interpretations~\cite{NaamanNRC2019,Hedegard2026}. Comparing the direct and reciprocal nonlocal signals measured with the same probe tests whether the junction is nonmagnetic, which is not established by the current-driven accumulations alone. Spin-to-charge conversion by chiral layers, the basis of the reciprocal measurement, has been observed by THz emission~\cite{Dong2025}. The constraints on $P_t$ and $P_r$ and the requirement of spin-flip reflection assume elastic transport. The relative sign of the two accumulations follows from symmetry alone and holds for any transport mechanism.

The constraints of Table~\ref{tab:classes} require equivalent contacts. Inequivalent contacts, as in a monolayer contacted by a tip, a single FM electrode, or defects~\cite{Dednam2023}, remove the lead-exchanging operation and lift the corresponding constraint, so that either class can be realized with any molecule. If these perturbations are sufficiently weak, the class is preserved except perhaps near energies where the polarization is small.

\begin{acknowledgments}
The code used to generate the data is available on Zenodo at 10.5281/zenodo.23425392. This work was supported by NSF award ECCS-2448290.
\end{acknowledgments}

\bibliographystyle{apsrev4-2}
\bibliography{SYM}

\end{document}